\documentclass[11pt]{article}

\usepackage{amsmath, amssymb, latexsym}

\newtheorem{thm}{\bf Theorem}[section]

\newtheorem{lem}[thm]{\bf Lemma}        
\newtheorem{prop}[thm]{\bf Proposition}  
\newtheorem{cor}[thm]{\bf Corollary}

\newtheorem{deff}[thm]{\bf Definition}

\newtheorem{expl}[thm]{\bf Example}

\newtheorem{remark}[thm]{\bf Remark}

\newcommand{\Bbox}{
{\unskip\nobreak\hfil\penalty50
\hskip1em\hbox{}\nobreak\hfil{\lower .5pt \hbox{$\Box$}}
\parfillskip=0pt \finalhyphendemerits=0 \par}
}

\newcommand{\eop}{
\ifmmode {\hbox{\Bbox}} \else \Bbox \fi
}

\newcommand{\ooo}{\omega ^{\omega^\omega }}
\newcommand\x{\times}

\newcommand{\N}{\mathbb{N}}
\newcommand{\Z}{\mathbb{Z}}
\newcommand{\lex}{<_\ell}   
\newcommand{\slex}{<_\ell}

\newcommand{\dom}{\textsf{dom}}

\newcommand{\A}{\mathcal{A}}
\newcommand{\Fr}{\mathsf{Fr}}
\newcommand{\Lfr}{\textsf{Lfr}}
\newcommand{\PBr}{\textsf{Pbr}}
\renewcommand{\L}{\mathcal{L}}

\newcommand{\F}{\mathcal{F}}

\newcommand{\Lin}{\mathbf{Lin}}
\newcommand{\zos}{\{0,1\}^*}
\newcommand{\boo}{\mathbf{1}}
\newcommand{\bfo}{\mathbf{1}}

\newcommand{\ol}[1]{\overline{#1}}
\title{Algebraic Linear Orderings}
\author{
S. L. Bloom\\
Department of Computer Science\\
Stevens Institute of Technology\\
Hoboken, NJ, USA\\
\and
Z. \'Esik\thanks{Supported in part by grant no. K 75249 from the 
National Foundation of Hungary for Scientific Research.} \\
Department of Informatics \\
University of Szeged\\
Szeged, Hungary}

\begin{document}
\maketitle

\begin{abstract}
An algebraic linear ordering is a component of the initial solution of a first-order 
recursion scheme over the continuous categorical algebra of countable linear orderings 
equipped with the sum operation and the constant $\boo$. Due to a 
general Mezei-Wright type result, algebraic linear orderings  are exactly
those isomorphic to the linear ordering of the leaves of an algebraic tree.
Using Courcelle's characterization of algebraic trees, 
we obtain the fact that a linear ordering is 
algebraic if and only if 
 it can be represented as the lexicographic ordering
of a deterministic context-free language. 
When the algebraic linear ordering is a well-ordering, its order type is an
algebraic ordinal. We prove that the Hausdorff rank of
any scattered algebraic linear ordering is less than $\omega^\omega$.
It follows that the  algebraic ordinals are exactly those less than 
$\omega^{\omega^\omega}$. 
\end{abstract}

\section{Introduction}


Fixed points and finite systems of fixed point equations, also called 
recursion schemes, occur in just about 
all areas of computer science.  For example, regular and context-free languages, 
rational and algebraic formal power series, finite state process 
behaviors can all be characterized as (components of) canonical solutions 
(e.g.,  unique, least or greatest, or initial solutions) 
of systems of fixed point equations.   
In this paper, we are interested in the solutions of systems
of fixed point equations over (countable) linear orderings.

Consider the fixed point equation
\begin{eqnarray*} 
X &=& \boo + X
\end{eqnarray*} 
over linear orderings, where $+$ denotes the sum operation (functor)
on linear orderings, and $\boo$ denotes a one-point linear ordering.
As it will be explained in the paper, 
its canonical solution is the ordinal $\omega$ 
(or any linear ordering isomorphic to  the ordering of the natural numbers). 
For another example, consider the system of fixed point equations 
\begin{eqnarray*}
X &=& Y + X\\
Y &=& \boo + Y
\end{eqnarray*}
The first component of its canonical solution is $\omega^2$,
and the second component is $\omega$. 
The canonical solution  
of  
\begin{eqnarray*}
X &=& X + \boo + X 
\end{eqnarray*} 
is the ordered set of the rationals.

As an example involving functors with parameters, consider
the system
\begin{eqnarray*}
F_0 &=& G(\boo)\\
G(x) &=& x + G(F(x))\\
F(x) &=& x + F(x)
\end{eqnarray*} 
The ordinal $\omega^\omega $ is the first component
of the canonical solution of this system.

We call a linear ordering \textbf{algebraic} if it is isomorphic to 
the first component of the canonical solution of a system of fixed point 
equations of the sort 
\begin{eqnarray}
\label{eq-sys}
F_i(x_0,\ldots,x_{n_i-1}) &=& t_i,\quad i = 1,\ldots, n
\end{eqnarray}
where $n_1 = 0$ and each $t_i$ is a term composed of the function variables 
$F_j$, $j = 1,\ldots, n$, the individual variables $x_0,\ldots,x_{n_i-1}$, 
the constant $\boo$ and the sum operation $+$.  (The meaning of ``canonical''
will be explained below.)  Moreover, we call a linear 
ordering \textbf{regular} if it is isomorphic to the first component of the canonical solution 
of a system (\ref{eq-sys}) with $n_i = 0$ for all $i$. 
Further, we call an ordinal $\alpha $ algebraic or regular\footnote{Our  
regular ordinals have nothing to do with the regular ordinals 
of set theory.} if it is
the order type of an algebraic or regular linear well-ordering.

It follows from the results in \cite{Courcelle78} and
\cite{BEmezei} that up to isomorphism, an algebraic (or regular) linear ordering 
is isomorphic to the leaf-ordering of 
the frontier of an algebraic (or regular) tree. In this way, we may represent 
algebraic and regular linear orderings and ordinals as frontiers of algebraic
or regular trees, and this is the approach we take here. 
Algebraic and regular trees were considered in \cite{CourcelleFund,Guessarian}.

By \cite{CourcelleFund}, a tree is regular (algebraic, respectively) iff 
its ``partial branch language'' is a regular language (deterministic context free language (dcfl),
respectively). Moreover, a ``locally finite'' tree is regular or algebraic
iff its ``branch language'' is regular, or a dcfl. We will use the representation 
of the algebraic linear orderings as frontiers of algebraic trees to derive
the result that a linear ordering is algebraic iff it is isomorphic to 
the lexicographic ordering of a dcfl (which additionally may be chosen to 
be a prefix language). A similar fact relating regular trees to regular languages
also holds.

The fact that the regular ordinals are exactly those less than $ \omega^\omega$
is immediate from the results of \cite{Heilbrunner}.
It was proved in \cite{BEord} that an ordinal is algebraic if and only if 
it is less than $ \omega^{\omega^\omega}$, see also \cite{BEbergen}.
In this paper, we will prove that the Hausdorff rank of any scattered linear ordering 
is less than $\omega ^\omega$, which extends one part of the result on algebraic
ordinals. 

We use ``prefix grammars'', introduced in \cite{BEord}, 
as our main tool in the analysis of
algebraic scattered linear orderings represented by tree frontiers.  
Scattered grammars defined below, are special cases of prefix grammars.
We show first that the lexicographic ordering of the context-free language 
generated by a scattered grammar has Hausdorff rank less than $\omega^\omega $.  
Then, we show that if the leaf ordering of an algebraic tree is scattered,
it is isomorphic to the lexicographic ordering of the language
generated by  a scattered grammar. This shows that all scattered algebraic
linear orderings are determined by scattered grammars.

This second result completes the 
proof that the Hausdorff rank of a scattered algebraic linear ordering
is less than $\omega^\omega$.  Since it is
easy to show that all ordinals less than $\ooo$ are
algebraic, we may conclude the result of \cite{BEord} that the algebraic
ordinals are precisely those less than $\ooo$.

The paper is organized as follows. In Section~\ref{sec-orderings}
we review several notions and results on linear orderings and establish 
a few simple facts for them that will be used in the sequel. In 
Section~\ref{sec-representation}, we use lexicographic orderings 
on (prefix) languages 
and leaf orderings of (binary) trees to represent all 
countable linear orderings. 
In Section~\ref{ssec: ccsa}, we recall the notions of continuous 
categorical algebras and present two examples: the algebras of trees 
and linear orderings. 
We also introduce recursion schemes and use them to define 
regular and algebraic elements (or objects) of continuous categorical algebras.
In particular, we obtain the notions of algebraic and regular trees 
and linear orderings. By a Mezei-Wright theorem, it follows 
that a linear ordering is algebraic iff it is isomorphic to the 
leaf ordering of an algebraic tree. Then in Section~\ref{sec-dcfl}, we prove that 
the algebraic linear orderings are exactly those linear orderings 
 isomorphic to the 
lexicographic ordering of a deterministic context-free (prefix) language. 
 Section~\ref{sec-closure} is devoted to closure 
operations on algebraic linear orderings. We derive some 
closure properties from the closure of algebraic trees 
with respect to 
substitution, 
while some additional closure properties come from the 
representation of algebraic linear orderings as lexicographic
orderings on deterministic context-free languages. 
In Section~\ref{sec-grammars}, we define 
prefix grammars and scattered grammars and show that the lexicographic 
ordering of the context-free language generated by a scattered grammar
has Hausdorff rank less than $\omega^\omega$. 
Then, in Section~\ref{sec-translation} we recall from \cite{BEord}  a translation 
of recursion schemes defining algebraic trees to prefix grammars. 
Relying on this translation, we prove
in Section~\ref{sec-completing} that the rank of every scattered algebraic linear ordering
is less than $\omega^\omega$. 
Section~\ref{sec-conclusion} contains some 
concluding remarks.


\section{Linear Orderings} 
\label{sec-orderings}

For the reader's convenience, in this section 
we have collected together some basic definitions and results
regarding linear orderings that will be used in the sequel. 
We start by recalling some basic definitions from 
\cite{Rosenstein}.

In this paper, a \textbf{linear ordering} $(P,<)$ is a \textit{countable} set 
$P$ equipped with a strict linear order relation $<$. 
\textit{Sometimes, we write just $P$ to denote a 
linear ordering $(P,<)$.}
(To force the collection of all linear orderings to be a small set, 
we may require that the underlying set of a linear ordering 
is a subset of a fixed set.)  
A \textbf{morphism} between 
linear orderings $(P,<) \to (Q,<)$ is a function $P \to Q$ 
which preserves the order relation (and is thus injective). 
The isomorphism class of a linear ordering is called an 
\textbf{order type}. Each \textbf{ordinal} may be identified 
with the isomorphism class of a \textbf{well-ordering}.
Since linear orderings in this paper are countable, 
so is each ordinal.
For basic facts for ordinals and ordinal arithmetic
we refer to \cite{Roitman} or \cite{Rosenstein}. 
We will sometimes identify isomorphic linear orderings.

\begin{deff}
Let $(P,<)$ be a linear ordering.
\begin{itemize}
  \item  A \textbf{subordering} of $P$ is a subset $Q$ of $P$ ordered by
the restriction of $<$ to $Q$. An \textbf{interval} of $P$ is a subordering
$I$ such that for all $x < y < z$ if $x,z \in I$ then $y \in I$. 
\item $P$ is \textbf{dense} if $P$ contains at least two points and
whenever $x<y$, there is some $z$ with $x<z<y$. (Note: in \cite{Rosenstein},
linear ordering containing zero or one point are also called dense.)
\item $P$ is \textbf{scattered} if $P$ has no subordering that is dense.
\end{itemize}
\end{deff} 

It is clear that every well-ordering is scattered. 
If $(Q_x,<_x)$ is a linear ordering,
for each point $x$ in the linear ordering $P$,
then the \textbf{$P$-sum} (``generalized sum'',
in the terminology of \cite{Rosenstein}), written
\begin{eqnarray*}
 \sum_{x \in P} (Q_x,<_x),
\end{eqnarray*}	
is the set $Q=\bigcup_{x \in P} Q_x \x \{x\}$ ordered as follows:
\begin{eqnarray*}
(y,x) < (y',x') &\iff & x<x' \text{ or } (x=x' \text{ and  } y<y').
\end{eqnarray*}
We mention two special cases. 
When  $P$ is a finite linear ordering, say  
$([n], <)$, where $[n] = \{0,1,\ldots, n-1\}$, we write
$\sum_{x \in P} (Q_x,<_x)$ as $Q_0 + \ldots + Q_{n-1}$.
And when for each $x\in P$, $Q_x$ is a fixed linear ordering $(Q,<)$, 
then $\sum_{x \in P} (Q_x,<_x)$ is isomorphic to the cartesian
\textbf{product} $Q \times P$.
By using generalized sum and product, we can define 
the \textbf{geometric sum} $\sum_{n \geq 0}P^n$,
where $P$ is any linear ordering.
For later use we also mention the \textbf{reverse} operation 
that takes a linear ordering $P = (P,<)$ to the linear ordering
$P^* = (P,>)$, where $x > y$ iff $y < x$, for all $x,y \in P$.

It is known, cf. \cite{Rosenstein}, that any scattered sum of scattered linear 
orderings is scattered. Moreover, the reverse of a scattered 
linear ordering is scattered. Hausdorff has classified scattered linear orderings 
according to their rank. Recall that for an ordinal $\alpha $,
the collection $V_ \alpha $ of linear orderings of Hausdorff rank
at most $ \alpha  $
is defined inductively as follows. $V_0$ consists of the empty linear ordering 
and the one-point linear orderings. 
Assuming that $V_ \beta $ is defined for all
ordinals $ \beta < \alpha $, $V_\alpha $ is defined as the collection
of all linear orderings
\begin{eqnarray*}
 \sum_{n \in \Z} (Q_n,<_n) 
\end{eqnarray*}
where $\Z$ is the linear ordering of the integers, 
and, for each $n \in \Z$, $Q_n \in \bigcup_{\beta<\alpha } V_{\beta }$.
(In \cite{Rosenstein}, the ``index set $\Z$'' was allowed to be
either a finite linear ordering,  or $ \omega $,
the linear ordering of the nonnegative integers, or
$ \omega^*$, the negative integers, or $\Z$. Since $V_0$ contains 
the empty linear ordering, the two definitions are equivalent.)

For a scattered linear ordering $Q$, the (Hausdorff) \textbf{rank of $Q$}, 
$r(Q)$, is the least ordinal $\alpha $ such that $Q \in V_ \alpha $.  
Hausdorff proved that every scattered linear ordering has a 
(countable) rank (see \cite{Rosenstein}). It is clear that for each 
ordinal $\alpha$, the collection of all linear orderings of rank 
$\alpha$ (or rank at most $\alpha$) is closed under the reverse operation. 

We will need some facts about the ranks of scattered linear orderings.
The first two facts are well-known, see  \cite{Rosenstein}. 

\begin{lem}
\label{lem: ordrank1}
If $\alpha $ is an ordinal, 
\begin{eqnarray*}
r( \omega^\alpha ) &=& \alpha .
\end{eqnarray*} 
\end{lem} 

\begin{lem}
\label{lem: ordrank2} 
If $P$ is a scattered linear ordering and $Q$ is a subordering
of $P$, then $Q$ is scattered with $r(Q) \leq r(P)$. In particular, if $\alpha,\beta$ are 
ordinals with $\alpha < \beta$, then $r(\alpha) \leq  r(\beta)$. 
\end{lem}

\begin{cor}
\label{cor: ordrank2}
Suppose that $\beta = \omega^\gamma$  such that $\gamma$ is a limit ordinal.
Then, for any ordinal $\alpha $,  $\alpha < \beta$ iff $r(\alpha) < \gamma$. 
\end{cor}

{\sl Proof.\/}
If $\alpha < \beta$ then since $\gamma$ is a limit ordinal, 
$\alpha < \omega^{\delta}$ for some $\delta < \gamma$. 
Thus, $r(\alpha) \leq r(\omega^\delta) = \delta < \gamma$.  
Conversely, if $r(\alpha) < \gamma = r(\beta)$, 
then it is not possible that $\beta \leq \alpha$, since 
otherwise we would have $r(\beta) \leq r(\alpha)$. 
Thus, $\alpha < \beta$. \eop

In particular, $\alpha < \omega^{\omega^n}$ for some ordinal $\alpha$ 
and positive integer $n$ iff $r(\alpha) < \omega^n$, and $\alpha < \ooo$ 
iff $r(\alpha) < \omega^\omega$.

We can use Lemma~\ref{lem: ordrank2} to establish the following 
fact for the rank of a finite sum of scattered linear orderings.  

\begin{cor}
 \label{sum-cor}
Suppose $A= B_1 + \ldots + B_k$, where each $B_i$ is
scattered, and suppose $\alpha=\max \{ r(B_i): 1 \leq i \leq k \}$.
Then $A$ is scattered and $\alpha \leq r(A) \leq \alpha + 1$.
\end{cor}  

{\sl Proof.\/ } The fact that $\alpha+ 1$ is an upper bound is clear from 
the definition of rank. Moreover, $\alpha$ is a lower bound by
Lemma~\ref{lem: ordrank2}. \eop

We now turn to the rank of generalized sums and products of scattered linear 
orderings. 

\begin{lem}
\label{lem-gen sum}
Suppose that $(P,<)$ is a scattered linear ordering of rank $\alpha$ 
and for $x \in P$, $(Q_x, <_x)$ is a scattered linear ordering
of rank at most $\beta$.  Then the $P$-sum
$Q  = \sum_{x \in P}(Q_x,<_x)$ is scattered of rank 
at most $\beta + \alpha$. 
\end{lem}

{\sl Proof.} For the fact that $Q$ is scattered, see \cite{Rosenstein}.
In order to prove that $r(Q) \leq \beta +\alpha$ we will argue by induction
on $\alpha$. When $\alpha = 0$ then $P$ is either empty or a singleton set
and our claim is obvious. Suppose that $\alpha > 0$. Then $P$ is isomorphic
to a generalized sum $\sum_{i \in \Z}(P_i,<_i)$, where each $(P_i,<_i)$ 
is scattered of rank less than $\alpha$.
Now for each fixed $i \in \Z$, 
the rank of $Q_i = \sum_{x \in P_i}(Q_x,<_x)$ is strictly less than $\beta+ \alpha$,
by the induction hypothesis. Since  $Q$ is isomorphic to the generalized sum
$\sum_{i \in \Z}Q_i$, it follows by the definition of the rank that
$r(Q) \leq \beta + \alpha$. \eop

\begin{cor}
\label{cor-product0}
If $P$ and $Q$ are scattered of rank $\alpha$ and $\beta$, respectively,
then $Q \times P$ is scattered of rank at most $\beta + \alpha$.
\end{cor}

{\sl Proof.\/ } Indeed, $Q\x P$ is isomorphic to $ \sum_{x \in P} Q_x$, 
where $Q_x=Q$, for each $x \in P$. \eop

\begin{cor}
If $P$ is a scattered linear ordering of rank $\alpha > 0$, then $\sum_{n\geq 0}P^n$ 
is scattered of rank at most $\alpha \times \omega$. 
\end{cor} 

{\sl Proof.\/ } For each $n\geq 0$, $r(P^n) \leq \alpha \times n < \alpha \times \omega$. 
Thus, by the definition of rank,  $r(\sum_{n\geq 0}P^n) \leq \alpha \times \omega.$ \eop 

In the sequel, we will deal with scattered linear orderings of rank $< \omega^n$,
for some positive integer $n$. In the next corollary, we give a summary of the 
above facts for such linear orderings.

\begin{cor}
\label{cor-product}
For every $n \geq 1$, the collection of all scattered linear orderings 
of rank less than $ \omega^n$ is closed under sum and cartesian product.
Moreover, if for each $k \in \Z$, $P_k$ is scattered with $r(P_k) < \omega^n$,
then $r(\sum_{k \in \Z}P_k) \leq \omega^n$. Finally, if 
$P$ is scattered with $r(P) < \omega^n$, then 
$r(\sum_{k \geq 0} P^k) \leq  \omega^{n}$. 
\end{cor}

\begin{cor}
\label{cor-ordinal closure}
For every $n \geq 1$, the collection of all ordinals 
less than $\omega^{\omega^n}$ is closed under sum and 
product. Moreover, if $\alpha  < \omega^{\omega^n}$
then $\alpha^\omega \leq \omega^{\omega^n}$. 
\end{cor}

{\sl Proof.\/} This follows from the previous corollary by noting that 
if $\alpha$ is a nonzero ordinal that is the order type of a linear 
ordering $P$, then the order type of $\sum_{k \geq 0}P^k$ is 
$\alpha^\omega$. \eop

We now give a characterization of the least collection of 
ordinals containing $0,1$ which is closed under the sum, 
product and $\omega $-power operations.

\begin{prop}
\label{closure prop}
The least collection of ordinals which contains 0,1, and is
closed under sum, product, and $\omega $-power
consists of the ordinals less than $\ooo$.  
\end{prop}
{\sl Proof.\/ }
The collection of ordinals less than $\ooo$ contains 0,1 and
is closed under sum,  product and $\omega $-power by
Corollary~\ref{cor-ordinal closure}. 

The converse follows from the Cantor Normal Form (see, e.g., \cite{Roitman}) 
for ordinals less than $\ooo$. 
\eop 

The rest of this section is 
devoted to unions of scattered 
linear orderings. 
The next result shows that the union of two scattered linear orderings is 
scattered.

\begin{lem}
\label{lem-union1}
Suppose that $P_0$ is a linear ordering, ordered by $<$, and $P,Q 
\subseteq P_0$. Then equipped with the restriction of the 
relation $<$ to $P$, $Q$ and $P \cup Q$, respectively, 
each of $P$, $Q$ and $P \cup Q$ is a linear ordering.
If $P$ and $Q$ are scattered, then so is $P \cup Q$.
\end{lem}

{\sl Proof.}
It is obvious that each of $P$, $Q$ and $P \cup Q$ is a linear ordering.
We prove only the last statement.
Assume to the contrary that $P\cup Q$ is not scattered, so that 
 $P \cup Q$ contains an infinite dense subordering $R$.
Since $R=(R\cap P) \cup (R\cap Q)$,
either $R \cap P$ or $R \cap Q$ contains at least two elements.
By symmetry, we may assume that $R \cap P$ does.
If $R \cap P$ is dense, then $P$ is not scattered,
contradicting the assumption.
If $R \cap P$ is not dense,
there are
$x_0<y_0$ in $R\cap P$ such that there is no $z \in R \cap P$ between them.
Then the set
 $ \{z \in R :x_0 < z < y_0\}$
is a subset of $Q$.  But this is a dense interval in $R$, contradicting 
the assumption that $Q$ is 
scattered. \eop 


The next goal in this section is to give an upper bound on the rank of $P \cup Q$, where 
$P$ and $Q$ are suborderings of a scattered linear ordering $P_0$. 
To formulate this result, we need a definition. 

\begin{deff}
For each ordinal $\alpha $, we define a positive integer
$m_\alpha $ as follows.  If $\alpha$ is $0$ or $\alpha$ is a limit ordinal,
$m_\alpha=1$.  If $\alpha=\beta +1$,
\begin{eqnarray*}
  m_\alpha &=& m_{\beta }+1.
\end{eqnarray*}
\end{deff}

Note: if $\alpha $ is a successor, $\alpha=\lambda + k+1$, 
where $\lambda $ is either 0 or a limit ordinal and $0 \leq k < \omega $,
and $m_\alpha = k + 2$. 
Thus, $1 \leq m_\alpha < \omega $, for all ordinals $\alpha $.

Below, in Proposition \ref{union upper bound prop}, we 
suppose that $P_0$ is a linear ordering,  and $P,Q 
\subseteq P_0$ are suborderings of $P_0$. 
We assume $P$ and $Q$ are scattered, 
so that  $P \cup Q$ is scattered also, by
Lemma \ref{lem-union1}.
Let 
$r(P) = \alpha$ and $r(Q) = \beta$.
We will prove the following upper bound on 
$r(P \cup Q)$.

\begin{prop}
\label{union upper bound prop}
  \begin{eqnarray*}
    r(P \cup Q) & \leq & \min \{\alpha + \beta + m_\beta , \beta + \alpha + m_\alpha \} .
  \end{eqnarray*}
\end{prop}

{\sl Proof.\/ } 
We will use induction on $r(P)$ to prove the following claim: 
For $P,Q$ as above, with 
$r(P) = \alpha$ and $r(Q) = \beta$, 
\begin{eqnarray*}
  r(P \cup Q) &\leq& \beta+ \alpha  + m_{ \alpha }.
\end{eqnarray*}

If $\alpha=0$, either $P=\emptyset  $ or $P$ is a singleton.
If $P= \emptyset  $, $P \cup Q=Q$, so $r(P \cup Q)= \beta = \beta+\alpha $
$< \beta+\alpha+1$.
If $P= \{p\} $, 
\begin{eqnarray*}
P \cup Q &=&  Q^{<p} + \{p \}  + Q^{>p}, 
\end{eqnarray*}
where $Q^{<p}=\{q \in Q: q<p\}$,  
$Q^{>p}=\{q \in Q: q>p\}$,  
so that, by Proposition \ref{sum-cor},
\begin{eqnarray*}
r(P \cup Q)& \leq & \max \{0,\beta  \} + 1 \\
& = &  \beta+ 1 \\
&=& \beta + \alpha + m_\alpha .
\end{eqnarray*}

Now assume $\alpha > 0$ and
\begin{eqnarray*}
  P &=& \sum_{n \in \Z} P_n,
\end{eqnarray*}
where the sets $P_n,\ n \in \Z$, are pairwise disjoint and $r(P_n)<\alpha $.
Define, for $n \in \Z$,
\begin{eqnarray*}
  Q_n &:=& \{q \in Q: \exists p \in P_n\, (p < q) \text{ and  } \exists p' \in P (q<p')
\text{ and  }  \forall m>n ( q<P_{m})\} .
\end{eqnarray*}
(Here, $q < P_m$ means that $q < p$ for all $p \in P_m$.)
Note: if $P_n=\emptyset  $, then $Q_n=\emptyset  $. Moreover, if $n<m$ and $q \in Q_n,\ q' \in Q_m$,
then $q<q'$. Also,
\begin{eqnarray*}
  P \cup Q &=& Q_{- \infty } + \sum_{n \in \Z} (P_n \cup Q_n) + Q_{\infty },
\end{eqnarray*}
where 
\begin{eqnarray*}
  Q_ {-\infty}  &:=& \{q  \in Q: q<P, \text{ i.e., } q<p, \text{ all }  p \in P\} \\
  Q_{\infty } &:=& \{q \in Q: P<q \} 
\end{eqnarray*}

Note that by Lemma~\ref{lem: ordrank2}, $r(Q_{- \infty}), r(Q_\infty ) \leq \beta $, and
 $ r(Q_n) \leq \beta $, for all $n \in \Z$.
Thus, by Proposition \ref{sum-cor} again,
\begin{eqnarray}
\label{eq:limit}
  r(P \cup Q) &\leq & \max \{\beta , r(\sum_{n \in \Z} P_n \cup Q_n )\} + 1.
\end{eqnarray}
Since $\alpha $ is countable,
there are two cases: either $\alpha $ is a successor or
$\alpha $ is the least upper bound of an increasing 
$\omega $-sequence.
\begin{enumerate}
\item If $\alpha $ is a successor, let $\alpha=\alpha_1+1$.  Then
$r(P_n) \leq \alpha_1$, for each $n$, and
by the induction hypothesis,
$r(P_n \cup Q_n) \leq \beta+\alpha_1+m_{\alpha_1} $. Thus,
\begin{eqnarray*}
  r(\sum_{n \in \Z} P_n \cup Q_n) & \leq & \beta+\alpha_1 + m_{\alpha_1}+1\\
&=& \beta+\alpha_1+ m_{\alpha}.
\end{eqnarray*}
So, by (\ref{eq:limit})
\begin{eqnarray*}
  r(P \cup Q) & \leq & \beta + \alpha_1 + m_\alpha + 1 \\
&=& \beta + \alpha + m_\alpha.
\end{eqnarray*}
\item Suppose $\alpha_1<\alpha_2< \ldots $ and $\alpha=\sup_k \alpha_k$.
Then, $r(P_n) \leq \alpha_{k_n}$, say, for each $n \in \Z$.
By the induction hypothesis,
$r(P_n \cup Q_n) \leq \beta+\alpha_{k_n}+m_{\alpha_{k_n}} <\beta+\alpha $. Thus,
\begin{eqnarray*}
  r(\sum_{n \in \Z} P_n \cup Q_n) & \leq & \beta+\alpha,
\end{eqnarray*}
so that by (\ref{eq:limit}),
\begin{eqnarray*}
  r(P \cup Q) &\leq & \beta + \alpha + 1 \\
&=& \beta+\alpha+m_\alpha . 
\end{eqnarray*}
\end{enumerate}
This ends the proof of the claim. 
Now, since the argument is symmetric, we have completed
the proof of Proposition~\ref{union upper bound prop}.   \eop

Since for any positive integer $n$, the ordinals less
than $\omega ^n$ are closed under addition, we obtain:

\begin{cor}
\label{cor-union}
Suppose that $P_0$ is a linear ordering and $P_1,\ldots,P_k
\subseteq P_0$. If $n \geq 1$ and each
$P_i$, $1 \leq i \leq k$ is scattered and has rank
less than $ \omega^n$, then $P_1 \cup \ldots \cup P_k$ 
is scattered of rank less than $ \omega^n$.  
\end{cor}

\section{Representation of Linear Orderings by Languages and Trees}  
\label{sec-representation}

In this section
we recall the fact that
that each linear ordering is isomorphic
to the lexicographic ordering of a (prefix) language over an ordered alphabet,
or the binary alphabet. Since lexicographic orderings of prefix languages
arise as frontiers of trees, it follows that every linear ordering
can be represented as the leaf ordering of a (binary) tree. 

Let the \textbf{alphabet} (i.e., finite nonempty set) $A$ be linearly ordered.
The set of \textbf{words}  on $A$, written  $A^* $, 
is equipped with two partial orders.  
The \textbf{prefix order}, written $u <_p v$,
is defined by:
\begin{eqnarray*}
  u <_p v &\iff & v = uw,
\end{eqnarray*}
for some nonempty word $w$.  
The \textbf{strict, or branching order}, written $u<_s v$, is defined
by:
\begin{eqnarray*}
  u <_s v &\iff & u=xav,\ v=x b w,
\end{eqnarray*}
for some words $x,v,w \in A^*$ and letters $a<b$ in $A$.
The \textbf{lexicographic order} on $A^*$ is defined by
\begin{eqnarray*}
  u \slex v &\iff & u<_p v \text{ or } u<_s v.
\end{eqnarray*}
It is easy to check that the lexicographic order is a linear order on $A^*$.

Recall from \cite{HopcroftUllman} that a language on $A$ is a subset of $A^*$.
A \textbf{prefix language}
on $A$ is a subset $L$ of $ A^*$ such that
if $u \in L$ and $uv \in L$ then $v$ is the empty word, written $\epsilon$.

Note that on a prefix language, the lexicographic order agrees with the 
strict order relation.

Since all linear orderings in this paper are countable, we 
may restrict attention to subsets of $\zos$ ordered by the 
lexicographic order, where $0 < 1$. 

\begin{prop}
\label{prop-langrep}
If $(P,<)$ is a countable linear ordering, there is a (prefix) language  $L \subseteq \zos$
such that $(P,<)$ is isomorphic to $(L,\lex)$.
\end{prop} 

{\sl Proof.\/ } This follows from the fact that, ordered by the lexicographic order,
the set of words $R$ denoted by the regular expression 
$(0 + 11)^*01$ (or $(00+11)^*01$)  is isomorphic to 
the rational numbers, ordered as usual. Further, any (countable) linear ordering may be 
embedded in the rationals. \eop

\begin{remark}
A linear ordering is called  \textbf{recursive} if it is isomorphic
to a linear ordering $(P,<)$ where  $P$ is a recursive subset of 
the set $\N$ of nonnegative integers, 
and the order relation $<$ is a recursive subset of $\N \times \N$. 
We show that each recursive linear ordering $(P,<)$ is isomorphic 
to a lexicographic ordering $(L,<_\ell)$ for some 
recursive (prefix) language $L \subseteq \zos$. Indeed, suppose that $P$ is a 
recursive subset of $\N$ and $<$ is a recursive subset of $\N \times \N$. 
When $P$ is finite our claim is clear, so we may assume that $P$ is 
infinite. Then let $p_0,p_1,\ldots$ be a recursive enumeration 
of the elements of $P$ without repetition. Consider the regular
language $R$ denoted by the regular expression $(0 + 11)^*01$.
As noted above, $(R,<_\ell)$ is isomorphic to the ordering
of the rationals. We define a morphism $h : P \to R$  as follows.
Suppose that for some $n \geq 0$, $h(p_0),\ldots,h(p_{n-1})$ have already 
been defined. Then let $h(p_n)$ be the lexicographically first element among
the shortest words 
$u$ of $R - \{h(p_0),\ldots,h(p_{n-1})\}$ such that for all $i = 0,\ldots,n-1$, 
$u <_\ell h(p_i)$ if and only if $p_n < p_i$. Let $L$ denote the set $h(R)$.
It is clear by construction that $L$ is recursive and 
$(P,<)$ is isomorphic to $(L,<_\ell)$.  
\end{remark}

We now turn to trees. Let $\Sigma$ be any \textbf{ranked alphabet}, so that $\Sigma$ 
is a finite
alphabet of the form $\Sigma = \bigcup_{n \geq 0} \Sigma_n$,
where $\Sigma_n$ is the set of letters (or function symbols) of rank 
(or arity) $n$. Let
$r(\Sigma)$, or just $r$  denote the largest integer $n$ such that 
$\Sigma_n$ is not empty. Recall that $[r] = \{0,\ldots,r-1\}$.  
Let $V = \{v_0,v_1,\ldots\}$ be 
a countable set of variables disjoint from $\Sigma$. 
One may define a (rooted, ordered) \textbf{tree} $T \in T^ \omega_\Sigma(V) $
as a partial function $[r]^* \to \Sigma \cup V$ whose domain, $\dom(T)$, is a 
prefix-closed subset of $[r]^*$ such that if $T(u) \in \Sigma_n$, 
and if $T(ui)$ is defined, then $i \in [n]$. Moreover, if 
$T(u) \in V$, then $T(ui)$ is not defined, for all $i \in [r]$. 
The words in the domain of $T$ are called the \textbf{vertices} of $T$. 
The \textbf{leaves} of $T$ are the words $u$ in the domain of $T$
such that $T(u) \in \Sigma_0 \cup V$.  The leaves of $T$ 
form a prefix language on $[r]$, denoted $\Fr(T)$. We let 
$T_\Sigma^\omega$ denote the collection of all trees 
$T \in T_\Sigma^\omega(V)$ such that $T(u) \in \Sigma$ 
for all vertices $u$. Note that 
$T_\Sigma^\omega$ contains the empty tree $\bot$. 
Each vertex $x$ of a tree $T \in T_\Sigma^\omega(V)$ 
is the  root of a \textbf{subtree} of $T$, denoted $T|_x$,
defined as follows: $T|_x(u) = T(xu)$ for all $u \in [r]^*$. 
A tree $T \in T_\Sigma^\omega(V)$ is called \textbf{finite} if its 
domain is finite and \textbf{complete} if it is not the empty tree
and whenever $T(u) \in \Sigma_n$ for some $n$, then 
$u\cdot 0,\ldots,u\cdot (n-1)$ are all in $\dom(T)$. 
Moreover, a  tree $T$ is \textbf{locally finite} if for each
vertex $x$ the subtree $T|_x$ contains at least one 
leaf. In particular, $\bot$ is locally finite.
The \textbf{frontier} or \textbf{leaf ordering} 
of a tree is the lexicographic ordering of its leaves. 


\emph{In the sequel, we will denote by $\Delta$ the ranked alphabet 
containing $2$ letters, $+$ and $\boo$, such that $+$ belongs to $ \Delta_2$ 
and $\boo$ belongs to $ \Delta_0$}. A tree in $T_\Delta^\omega$ 
will  be referred to as a \textbf{binary tree}. Below we
will usually write $t,s$ to denote finite trees and $T,S$ 
to denote possibly infinite trees.

All linear orderings are isomorphic to frontiers of 
binary trees. 
\begin{prop} 
\label{prop-frontier-1}
\cite{Courcelle, CourcelleFund} 
For any (countable)  linear ordering $P$, there is a (locally finite) 
binary tree $T$ such that $(\Fr(T),\lex)$, the leaf ordering of $T$, 
is isomorphic to $P$. 
\end{prop}  

{\sl Proof.\/} By Proposition~\ref{prop-langrep}, each linear ordering
is isomorphic to the lexicographic ordering $(L,\lex)$ of a 
prefix language $L \subseteq \zos$. But any such linear ordering
$(L,\lex)$ is the leaf ordering of some locally finite binary tree. \eop

\section{Continuous Categorical $\Sigma$-algebras and Recursion Schemes}
\label{ssec: ccsa}

In this section we review the notion of continuous categorical algebras
and recursion schemes over such algebras. Every recursion scheme 
has a canonical solution over any continuous categorical algebra giving
rise to algebraic and regular objects, or elements. Two special cases 
will be of crucial importance for this paper. When the 
algebra is the $\Delta $-algebra of trees or linear orderings, we obtain the 
notions of algebraic and regular trees and linear orderings. 

Suppose that $\Sigma = \bigcup_{n \geq 0}  \Sigma_n$ is a ranked alphabet.
A \textbf{categorical $\Sigma$-algebra}  
$\A$ consists of a (small) category, also denoted  $\A$, 
together with a functor  $\sigma^\A : \A^n \to \A$,   
for each letter $\sigma \in \Sigma_n$, $n \geq 0$, 
called the operation induced by $\sigma$.
A \textbf{morphism} of categorical $\Sigma$-algebras is a functor which preserves
the operations up to natural isomorphism.
(See \cite{BEregular,BEbergen,BEmezei}.) 

We say that a categorical $\Sigma$-algebra $\A$ is \textbf{continuous} if 
it has initial object and colimits of $\omega$-diagrams; moreover, 
the operations $\sigma^\A$ are continuous, i.e., they preserve colimits of 
$\omega$-diagrams in each argument.  \textbf{Morphisms} of continuous categorical 
$\Sigma$-algebras are continuous and preserve initial objects. 
 
The notion of continuous categorical $\Sigma$-algebra generalizes the 
notion of continuous ordered $\Sigma$-algebra \cite{ADJ,Guessarian}, 
where the underlying category is a poset. Two examples of continuous 
categorical $\Sigma$-algebras are given below. 

The set of trees $T_\Sigma^\omega(V)$ (or $T_\Sigma^\omega$) is turned into
a $\Sigma$-algebra in the usual way, cf. \cite{Guessarian,CourcelleFund,ADJ}.
 Moreover, for any two trees $T,S$,
we define $T < S$ iff $\dom(T)$ is a proper subset of $\dom(S)$ 
and for each $u \in \dom(T)$, $T(u) = S(u)$. It is known that
$T_\Sigma^\omega(V)$ and $T_\Sigma^\omega$ are continuous ordered 
$\Sigma$-algebras. Moreover, $T_\Sigma^\omega(V)$ is freely generated 
by $V$ in the class of all continuous categorical $\Sigma$-algebras:
For any function $h: V \to \A$ into a continuous categorical
$\Sigma$-algebra, there is up to isomorphism a unique morphism 
$h^\sharp$ of categorical $\Sigma$-algebras extending $h$.
In particular, $T_\Sigma^\omega$ is an initial continuous 
categorical $\Sigma$-algebra. See \cite{ADJ,Guessarian,BEmezei} 
for more details. 

The second example involves linear orderings.
The category $\Lin$ of linear orderings has linear orderings 
as its objects and order preserving maps as morphisms, see also 
Section~\ref{sec-orderings}. It has as initial object the empty linear 
ordering denoted $\mathbf{0}$. Moreover, $\Lin$ has colimits of 
all $\omega$-diagrams. 

We have already defined the sum of any two linear orderings. 
The sum operation can naturally be extended to a functor
$\Lin^2 \to \Lin$.  The sum $h_1 + h_2$ 
of morphisms $h_i: P_i \to Q_i$, $i = 1,2$  is defined so that 
it agrees with $h_i$ on $P_i$, for $i = 1,2$. By letting 
$\boo$ denote a singleton ordering, $\Lin$ becomes a continuous 
categorical $\Delta$-algebra. For more details, see \cite{BEmezei}.

\subsection{Recursion schemes}
   
Let $\Sigma $ be a ranked set. Recall from Section~\ref{sec-representation} 
the definition of $\Sigma$-trees $T^\omega_\Sigma(V)$ with
variables in the set $V = \{v_0,v_1,\ldots\}$. Each 
complete finite $\Sigma$-tree may be identified with a $\Sigma$-\textbf{term},
or term for short.

\begin{deff}
\label{def: recursion scheme}
A \textbf{recursion scheme} over $\Sigma$
is a sequence $E$ of equations
\begin{eqnarray}
\nonumber
F_1(v_0,\ldots,v_{k_1-1}) &=&  t_1 \\
\label{rec scheme}
 & \vdots & \\
\nonumber
 F_n(v_0,\ldots,v_{k_n-1}) &=& t_n
\end{eqnarray}
where $t_i$ is a term over the ranked alphabet $\Sigma \cup \F$ 
in the variables $v_0,\ldots,v_{k_i-1}$,
for $1 \leq i \leq n$, where $\F=\{ F_1, \ldots, F_n\}$. 
A recursion scheme is \textbf{regular} if $k_i=0$, for 
\textit{each} $i$ with $1 \leq i \leq n$.
\end{deff}
In the above definition, $\Sigma \cup \F$ is the ranked alphabet whose 
letters are the letters in $\Sigma$ together with the letters in 
$\{F_1,\ldots,F_n\}$ where each $F_i$ is of rank $k_i$. The 
letters $F_i$ are called function (or functor) variables. 

In any continuous categorical $\Sigma$-algebra $\A$, any 
scheme $E$ as in 
(\ref{rec scheme})
 induces a continuous endofunctor $E^\A$ over the 
category 
$$[\A^{k_1} \to \A] \times \ldots [\A^{k_n} \to \A],$$
where $[\A^k \to \A]$ denotes the category of all 
continuous functors $\A^k \to \A$. Since this category
also has initial object and colimits of $\omega$-diagrams,
it has an \textbf{initial fixed point} $|E^\A|$ 
which is unique up to isomorphism (See  \cite{Adamek,Wand}).

\begin{deff}
Suppose that $\A$ is a continuous categorical $\Sigma$-algebra. 
We call a functor $f: \A^m \to \A$,
\textbf{algebraic} if there is a recursion scheme $E$
such that $f$  is isomorphic to $|E|_1^A$, the first component 
of the above initial solution. When $m = 0$, $f$ may be 
identified with an object of $A$, called an \textbf{algebraic object},
or \textbf{algebraic element}. 
An object (or element) $a$ in $\A$ is \textbf{regular} if
there is a regular recursion scheme $E$
such that $a$ is isomorphic to $|E|_1^\A$. 
\end{deff}

By applying the above notion to $\Lin$ or $T_\Sigma^\omega(V)$ or $T_\Sigma^\omega$, we obtain the 
notions of algebraic and regular linear orderings, 
and algebraic and regular trees, respectively. 

\begin{deff}
For any ranked alphabet $\Sigma$, 
we call a tree $T\in T_\Sigma^\omega(V)$ (or $T \in T_\Sigma^\omega$) 
an \textbf{algebraic tree}
(\textbf{regular tree}, respectively) if it is an algebraic element 
(regular element, respectively) of 
the continuous categorical $\Sigma$-algebra $T_\Sigma^\omega(V)$ (or $T_\Sigma^\omega$).
We call a linear ordering $(P,<)$ an \textbf{algebraic linear ordering}
(\textbf{regular linear ordering}, respectively) if it is an 
algebraic object  (regular object, respectively) of 
the continuous categorical $\Delta$-algebra $\Lin$.
\end{deff}

By a general Mezei-Wright theorem \cite{BEmezei}, morphisms between
continuous categorical algebras preserve algebraic and regular 
objects. Since $T_\Delta^\omega$ is initial continuous categorical
$\Delta$-algebra, up to natural isomorphism there is a unique 
morphism $T_\Delta^\omega \to \Lin$. This essentially unique 
morphism maps a tree to its leaf ordering. So we obtain:

\begin{prop}
\label{prop-frontier}
\cite{BEmezei}
A linear ordering is algebraic or regular if and only if it is isomorphic
to the frontier of an algebraic or regular tree in $T_{\Delta}^\omega$.
\eop 
\end{prop} 

(For the case of regular trees and regular linear orderings see 
also \cite{Courcelle78}.)

The leaf ordering of an algebraic tree over \textit{any} 
ranked alphabet $\Sigma$ is an algebraic linear ordering.
\begin{prop}
For any ranked alphabet $\Sigma $ and any algebraic tree $T\in T_\Sigma(V)$ 
there is an algebraic
tree $T'\in T_\Delta^\omega$ such that $(\Fr(T),\lex)$ and $(\Fr(T'),\lex)$ are isomorphic. \eop 
\end{prop}

For example, consider the system $E$ of equations
\begin{eqnarray*}
F_1 &=& \sigma_3(a,b,F_2(a))\\ 
F_2(x) &=& F_3(x,x)\\ 
F_3(x,y) &=& \sigma_3( \sigma_1(a),\ F_3(x,\ F_3(x,y)),\  y)
\end{eqnarray*} 
which involves a function symbol $\sigma_3 $ in $\Sigma_3$
and a function symbol $\sigma_1$ in $\Sigma_1$.   The least
solution consists of three trees $(T_1,T_2,T_3)$ having vertices 
of out-degree  3 and 1, labeled $\sigma_3$ and $\sigma_1$. 
We replace the system $E$ by the system
\begin{eqnarray*}
F_1 &=& +(\bfo ,\ +(\bfo ,F_2(\bfo )))\\ 
F_2(x) &=& F_3(x,x)\\
F_3(x,y) &=& + (\bfo ,\ +( F_3(x,\ F_3(x,y)),\ y) )
\end{eqnarray*} 
in which the right hand terms involve only the function variables and the one
binary function symbol $+ $, and the one constant symbol $\bfo $.
If $(T'_1, T'_2, T'_3)$ is the least solution of this second
system, 
$  (\Fr(T_i),\lex) $ is isomorphic to $(\Fr(T'_i), \lex)$, for $i = 1,2,3$.

Several characterizations  of algebraic and regular trees can be found in 
\cite{ADJ,CourcelleFund,Guessarian}. 
For characterizations of regular linear orderings we refer to \cite{Courcelle,BEs}. 
See also Section~\ref{sec-dcfl}.

In Section \ref{sec-closure}, we will make use of the tree $T_0$ in Example 
\ref{example-omega}.
\begin{expl}
\label{example-omega}
Let $\Sigma$ contain the binary symbol $+$, the unary symbol $f$ and the constant $\bfo$.
Consider the system
\begin{eqnarray*}
F_0 &=& F(\bfo)\\
F(x) &=& + (x, F(f(x)))
\end{eqnarray*} 
Then the first component of the least solution of this system is 
the tree 
$$T_0 = +(\bfo ,+(f(\bfo ),+(f(f(\bfo )), \ldots ,+(f^n(\bfo ),\ldots )))).$$
Thus, this tree is algebraic. See also \cite{Guessarian}, p. 39. 
\end{expl}

\subsection{Substitution on trees}

We will derive some closure properties of algebraic linear orderings 
from the closure of algebraic trees with respect to substitution.

Suppose that for each $\sigma \in \Sigma_n$ we are given 
a tree $S_\sigma \in T^\omega_\Gamma(V_n)$, i.e., a tree 
in $T_\Gamma^\omega(V)$ such all leaves labeled in $V$ 
are actually labeled in the set $\{v_0,\ldots,v_{n-1}\}$.  For each finite 
tree $t \in T_\Sigma(V)$ we define a tree $R = t[\sigma \mapsto S_\sigma]_{\sigma \in \Sigma}$,
sometimes denoted just $t[\sigma \mapsto S_\sigma]$
by induction on the size of $t$. When $t$ is the empty tree $\bot$, 
so is $R$. When $t$ is $x$, for some $x \in V$, then $R = x$.
Otherwise $t$ is of the form $\sigma(t_0,\ldots,t_{m-1})$, and we define 
$$R = S_\sigma[v_0 \mapsto S_1,\ldots,v_{m-1} \mapsto S_{m-1}]$$  
where $S_i = t_i[\sigma \mapsto S_\sigma]$, for all $i$. 

Suppose now that $T$ is an infinite tree in $T_\Sigma(V)$. Then there is an 
ascending $\omega$-chain $(t_n)_n$ of finite trees such that 
$T = \sup_n t_n$. We define 
$$T[\sigma \mapsto S_\sigma] = \sup_n t_n[\sigma \mapsto S_\sigma].$$

The following facts are known, see \cite{CourcelleFund}.
\begin{prop}
\label{proposition-subst}
Substitution is a continuous function
$$T^\omega_\Sigma(V) \times \prod_n  T^\omega_\Gamma(V_n)^{\Sigma_n} \to T^\omega_\Gamma(V).$$ 
Further, the classes of algebraic and regular trees are closed
under substitution.
\end{prop}

\section{Algebraic Linear Orderings and Deter\-min\-istic Con\-text-free 
Languages}
\label{sec-dcfl}

In this section we define context-free and deterministic context-free
linear orderings using lexicographic orderings on cfl's and dcfl's,
i.e., context-free and deterministic context-free languages. The
main result of this section shows that a linear ordering is algebraic
iff it is deterministic context-free. This result follows 
easily from Courcelle's characterization of algebraic trees 
by dcfl's \cite{Courcelle}. While every deterministic
context-free linear ordering is context-free, it remains open
whether there is a context-free linear ordering that is not 
deterministic context-free.

\begin{deff}
A linear ordering is \textbf{context-free} 
(\textbf{deterministic context-free}, respectively) 
if it is isomorphic to the lexicographic ordering 
of a cfl (dcfl, respectively) over some (ordered) alphabet 
$A$ (or equivalently, over the $2$-letter alphabet $\{0,1\}$). 
\end{deff}

As the next proposition shows, we may restrict ourselves to
cfl's or dcfl's which are prefix languages.

\begin{prop}
\label{prop-prefix}
A linear ordering is context-free
(deterministic context-free, respectively) 
iff it is isomorphic to a linear ordering $(L,\lex)$ 
for some context-free (deterministic context-free, respectively) 
\textbf{prefix} language for some (ordered) alphabet 
$A$ (or equivalently, over the $2$-letter alphabet $\{0,1\}$).  
\end{prop}

{\sl Proof.} We only prove this fact for the two-letter alphabet 
$\{0,1\}$. Let $L \subseteq \zos$ be a context free language.
If $L$ is not a prefix language, then let $L' = L(-1)$, a context-free
language on the three-letter alphabet $\{-1,0,1\}$, ordered by 
$-1 < 0 < 1$. It is clear that $L'$ is a context-free prefix language
and $(L',\lex)$ is isomorphic to $(L,\lex)$. To end the proof, let
us introduce the following encoding of the three letter alphabet
$\{-1,0,1\}$: $h(-1) = 00$, $h(0) = 01$ and $h(1) = 10$.
Then $h$ extends to an (injective) homomorphism $\{-1,0,1\}^* 
\to \zos$ as usual, and $(L,\lex)$  is isomorphic to 
$(h(L'),\lex)$. Moreover, when $L$ is deterministic, so is 
$h(L')$.
\eop

Each tree can be represented in several
ways by languages. Recall that a tree $T \in T_\Sigma^\omega$ 
is nothing but a partial  function $[r]^* \to \Sigma$, where $r = r(\Sigma)$, 
subject to certain properties.  Defining a function $[r]^* \to \Sigma$
amounts to specifying the languages $T^{-1}(\sigma)$, for all 
$\sigma\in \Sigma$, or a single language which is a combination
of these languages. Below we review from \cite{Courcelle78,CourcelleFund} a possible
representation of trees by (partial) branch languages. 

Given $\Sigma$, we introduce the (unranked) alphabet 
$\ol{\Sigma} = \{(\sigma,i) : i\in [n],
\ \sigma \in \Sigma_n,\ n > 0\}$. When $T$ 
is a tree in $T_\Sigma^\omega$ and $u \in \dom(T)$, we define 
$\widehat{u}\in \ol{\Sigma}^*$ by induction.
First, $\widehat{\epsilon } = \epsilon$.  
Next, when $u = vi$ with  $T(v) = \sigma$ and $i \in [r]$,  
then  $\widehat{u} = \widehat{v}(\sigma,i)$. 
It is clear that for any $T$, the 
\textbf{partial branch language} 
\begin{eqnarray*}
\PBr(T) &=& \{\widehat{u}T(u) : u \in \dom(T)\}
\end{eqnarray*}
over the alphabet $\Sigma \cup \ol{\Sigma}$ completely describes $T$.
We also define the \textbf{branch language} or 
\textbf{labeled frontier language} of $T$ as the set 
\begin{eqnarray*}
\Lfr(T) = \{\widehat{u}T(u) : T(u) \in \Sigma _0\}. 
\end{eqnarray*}

For the following result see \cite{CourcelleFund}, Theorem 5.5.1. 
\begin{thm} 
\label{thm-Courcelle alg}
Suppose that $T\in T_\Sigma^\omega$ is a \textbf{complete} tree.
If  $T\in T_\Sigma^\omega$ is algebraic 
then $\PBr(T)$ and $\Lfr(T)$ are dcfl's. Moreover, 
if $\PBr(T)$ is a dcfl or when $T$ is locally finite and $\Lfr(T)$ 
is a dcfl, then $T$ is algebraic. 
\end{thm}

In order to make the above result applicable to trees in $T_\Sigma^\omega$ 
that are not necessarily
complete, we need to describe a procedure for completing trees.  
Let $\Omega$ be a new letter of rank $0$ and let $\Sigma_\Omega$ denote the
ranked alphabet obtained by adding the letter $\Omega$ to $\Sigma$. Then
each tree $T \in T_\Sigma^\omega$ has a \textbf{completion} $T_\Omega \in T_{\Sigma_\Omega}^\omega$ 
defined as follows:
For all $u \in [r]^*$ where $r = r(\Sigma)$, if $T(u)$ is defined then $T_\Omega(u) = T(u)$.
If $T(u)$ is not defined but $u = \epsilon$ or $u = vi$ for some $v$ 
and $i \in [n]$ such that $T(v) \in \Sigma_n$, then we let  
$T_\Omega(u) = \Omega$.

The next fact is clear: 

\begin{lem}
\label{lem-completion}
For any tree $T \in T_\Sigma^\omega$, $T$ is algebraic iff $T_\Omega$ is.
\end{lem}

By Lemma~\ref{lem-completion} and Theorem~\ref{thm-Courcelle alg} 
we immediately have: 

\begin{cor}
Suppose that $T \in T_\Sigma^\omega$. If $T$ is algebraic
then $\PBr(T_\Omega)$ and $\Lfr(T_\Omega)$ are dcfl's. Moreover, 
if $\PBr(T_\Omega)$ is a dcfl or if $T_\Omega$ is locally finite
and $\Lfr(T_\Omega)$ is a dcfl, then $T$ is algebraic. 
\end{cor}

Using the above corollary, we now establish the following
(slight) generalization of Theorem~\ref{thm-Courcelle alg}.

\begin{prop}
Suppose that  $T \in T_\Sigma^\omega$. If $T$ is algebraic 
then $\PBr(T)$ and $\Lfr(T)$ are dcfl's. Moreover,
if $\PBr(T)$ is a dcfl or if $\Lfr(T)$ is a dcfl and $T$
is locally finite, then $T$ is algebraic. 
\end{prop}

{\sl Proof.} Suppose first that $T$ is algebraic. Then 
$T_\Omega$ is also algebraic and thus $\PBr(T_\Omega)$ 
and $\Lfr(T_\Omega)$ are dcfl's as are 
$\PBr(T)= \PBr(T_\Omega) \cap \ol{\Sigma}^*\Sigma$ and $\Lfr(T) = 
\Lfr(T_\Omega) \cap \ol{\Sigma}^*\Sigma_0$. Here we have
used the well-known fact that the intersection of a dcfl with a regular
language is a dcfl. 

Suppose now that $\PBr(T)$ is a dcfl and consider a deterministic 
pushdown automaton (dpda) accepting $\PBr(T)$ with final states,
see \cite{HopcroftUllman}.
We modify this dpda by adding to the set of states a new state 
$q_\Omega$ and a new accepting state $s_\Omega$,  
and by adding rules ensuring that whenever the dpda is able 
to move from a configuration $c$ to an accepting configuration 
while reading a letter $\sigma \in \Sigma_n$,
$n > 0$, but at the same time $c$ has no successor configuration 
with respect to the letter $(\sigma,i)$ for some $i$, then 
the new dpda will be able to move first to state $q_\Omega$
while reading $(\sigma,i)$ and then to $s_\Omega$ 
while reading $\Omega$. The resulting dpda accepts $\PBr(T_\Omega)$.
Thus, by Theorem~\ref{thm-Courcelle alg} and Lemma~\ref{lem-completion}, 
$T_\Omega$ and $T$ are algebraic. 

Suppose finally that $T$ is locally finite and $L=\Lfr(T)$ 
is a dcfl. A similar but somewhat more involved construction 
works to show that $\Lfr(T_\Omega)$ is a dcfl. Indeed, 
there exists a dpda accepting $L$ such that whenever 
from the initial configuration  the dpda can reach a 
configuration  while reading a word 
$u$, then there is some word $v$ such that $uv \in \Lfr(T)$. 
This holds because $\Lfr(T)$ is not empty and since 
the dpda obtained from the canonical LR(1) parser 
\cite{AhoUllman} has this property. 
Then we proceed as above. We add two states $q_\Omega$ and $s_\Omega$ 
and new transitions to the effect that if a given
configuration $c$ has a successor configuration 
for a letter $(\sigma,i)$ but at the same time $c$ has 
no successor configuration for $(\sigma,j)$, for some $j$, 
then the new dpda is able to move from $c$ to state $q_\Omega$ while reading $(\sigma,j)$ 
and then to the state $s_\Omega$ while reading the letter $\Omega$. \eop

We now consider 
trees over the ranked alphabet $\Delta$ defined above.
In this case we can identify the letter $(+,0)$ with $0$ and the letter 
$(+,1)$ with 1, so that for any tree $T$, $\PBr(T)$ may be viewed as a 
subset of $\zos \Delta$ and $\Lfr(T)$ as a subset of $\zos \boo$.
The main contribution of this section is:

\begin{thm}
A linear ordering is algebraic iff it is deterministic context-free.
\end{thm}

{\sl Proof.} Suppose that $(P,<)$ is an algebraic linear ordering.
Then there is an algebraic tree $T \in T_\Delta^\omega$ such that 
$(P,<)$ is isomorphic to $(\Fr(T), \lex)$. But $\Fr(T)$ is the right quotient
of the deterministic context-free language $\Lfr(T)$ with respect to 
the letter $\boo$ and is thus also a dcfl. (See \cite{HopcroftUllman}.) 
We conclude that $(P,<)$ is 
a deterministic context-free linear ordering.

Suppose now that $(P,<)$ is isomorphic to $(L,\lex)$ where 
$L \subseteq \zos$ is a dcfl. By Proposition~\ref{prop-prefix}, 
we may suppose that $L$ is a 
prefix language. 
Define the tree $T \in T_\Delta^\omega$ by 
\begin{eqnarray*}
T(u) &=& 
\left\{
\begin{array}{ll}
+ & {\rm if}\ uv \in L \ {\rm for}\ {\rm some}\ v \neq \epsilon\\
\boo & {\rm if}\ u \in L\\
{\rm undefined} & {\rm otherwise},
\end{array}
\right.
\end{eqnarray*} 
for all $u \in \zos$.
Then $T$ is a locally finite tree with $\Fr(T) = L$. 
Since $\Fr(T)$ is a dcfl, so is $\Lfr(T) = \Fr(T)\boo$.
Thus, since $T$ is locally finite, $T$ is algebraic 
proving that $(P,<)$ is also algebraic. \eop

\begin{cor}
A linear ordering $(P,<)$ is algebraic iff $P$ is empty or $(P,<)$  is isomorphic to 
$(\Fr(T), \lex)$ for some locally finite algebraic tree $T \in T_\Delta^\omega$. 
\end{cor}

{\bf Open Problem}. Does there exist a context-free linear ordering that is not a 
deterministic context-free linear ordering? 
\

\section{Closure Properties of Algebraic Linear Orderings}
\label{sec-closure}

In this section, we establish some closure properties of algebraic
linear orderings. In particular, we will show that algebraic linear 
orderings are closed under finite sum and product as well as 
infinite geometric sum. 
These closure properties will readily imply that all ordinals less than $\ooo$ 
are algebraic.

We will derive the closure of algebraic linear orderings under 
sum, product and infinite geometric sum from the closure of 
algebraic trees under 
substitution. 
We start with sum and product. Recall the ranked alphabet $\Delta $
defined in Section \ref{ssec: ccsa}.

The next three Propositions show that algebraic linear orderings are 
closed under sum, product, and geometric sum. They were proved in \cite{BEord} for 
algebraic well-orderings. The same arguments work in the slighly 
more general setting of algebraic linear orderings. 
We repeat them for the reader's convenience. (The reader is
invited to prove the same closure properties using 
deterministic context-free languages.) 
Assume that $P$ and $Q$ are respectively the leaf orderings
of the  trees $T,S$ in $T^\omega_\Delta$.
\begin{prop}
\label{sum prop}
If $P$ and $Q$ are algebraic linear orderings, so is $P + Q$.
\end{prop}

 {\sl Proof.\/ }   Consider the  tree $+(a,b)$,
 where $a,b$ are letters of rank $0$. Then the frontier of the 
 tree $+(T,S) = +(a,b)[a \mapsto T,b\mapsto S]$ is isomorphic to $P + Q$.
 Moreover, when $T,S$ are algebraic, then so is $+(T,S)$.
  \eop

 \begin{prop}
 \label{product prop}
 If $P$ and $Q$ are algebraic linear orderings, so is $Q \times P $.
 \end{prop}

 {\sl Proof.\/ } 
  The frontier of the tree $T[\bfo \mapsto S]$ obtained by substituting 
  a copy of $S$ for each leaf of $T$ isomorphic to  $Q \times P$;
  it is algebraic if $T$ and $S$ are. 
  \eop 

\begin{prop}
\label{omega pow prop}
If $P$ is an algebraic linear ordering, then so is $\sum_{n\geq 0}P^n$.
\end{prop}
 {\sl Proof.\/ }  Let $T'$ result from $T$ 
 by relabeling each leaf by the variable $x_0$. Consider the tree $T_0$
 of Example~\ref{example-omega} and let $S$ be the  tree obtained
 by substituting the tree $T'$ for each vertex labeled $f$:
 $S = T_0[f \mapsto T']$. 
 Then the frontier of $S$ is isomorphic to  $\sum_{n\geq 0}P^n$. Moreover,
 if $T$ is algebraic then so is $S$. 
 \eop 
Since for any (countable) ordinal $\alpha > 1$, the ordinal $\alpha^\omega$  
is $\sum_{n\geq 0}\alpha^n$, we immediately obtain: 

\begin{cor}
\label{cor-closure prop2}
The collection of algebraic ordinals is
closed under sum, product, and $\omega $-power. 
\end{cor}

\begin{cor}
\label{cor: ub on ords}
Every ordinal less than $\ooo$ is algebraic. 
\end{cor}

{\sl Proof.\/ }
The ordinal 0 is algebraic, since if $t$ is the empty tree in 
$T_\Delta^\omega$, then $\Fr(t)$ is the empty language whose frontier 
represents the ordinal $0$. 
The ordinal 1 is algebraic since the tree $\bfo$ is algebraic. 
Thus the result follows from Proposition~\ref{closure prop} and the 
closure properties of algebraic ordinals, Corollary~\ref{cor-closure prop2}. \eop 

For the facts in Corollary~\ref{cor-closure prop2} and Corollary~\ref{cor: ub on ords}, 
see also \cite{BEbergen}.

We establish two more closure properties of algebraic linear orderings.
To prove these results, we will rely on the characterization of algebraic 
linear orderings as lexicographic orderings of dcfl's.. 

\begin{prop}
\label{interval-prop}
If $P$ is an algebraic linear ordering and $I$ is an interval of $P$, 
then $I$ is algebraic. 
\end{prop} 
{\sl Proof.\/ } Since $P$ is algebraic, it can be represented 
as the lexicographic ordering of a deterministic context-free 
language $L \subseteq \{ 0,1 \}^*$. If  $y_0 \in \{0,1\}^*$, then 
consider the languages 
\begin{eqnarray*} 
R_{<_{\ell} u_0} &=& \{u \in \{0,1\}^* : u <_{\ell} y_0\}\\
R_{\leq_{\ell} u_0} &=& \{u \in \{0,1\}^* : u \leq_{\ell} y_0\}\\
R_{>_{\ell} u_0} &=& \{u \in \{0,1\}^* : u >_{\ell} y_0\}\\
R_{\geq_{\ell} u_0} &=& \{u \in \{0,1\}^* : u \geq_{\ell} y_0\}.
\end{eqnarray*} 
Clearly, all of them are regular. Since $I$ can be represented 
as the lexicographic ordering of a finite intersection of $L$
with such languages, and since the intersection of a 
deterministic context-free language with a regular language 
is deterministic context-free, it follows that $I$ is  algebraic. 
\eop

\begin{remark}
By Proposition \ref{interval-prop}, 
it follows that whenever $\alpha$ is an algebraic ordinal, 
then so is any ordinal less than $\alpha$. Using this fact, we can give an alternative
proof of Corollary~\ref{cor: ub on ords}. Indeed, as noted above, all ordinals of the 
form $\omega^{\omega^n}$ are algebraic, for $n \geq 0$. Since each 
ordinal less than $\ooo$ is less than $\omega^{\omega^n}$, for some $n$, 
all ordinals less than $\ooo$ are algebraic.  
\end{remark}

\begin{prop}
The class of algebraic linear orderings is closed under the reverse 
operation. 
\end{prop}

{\sl Proof.\/} Suppose that $P$ is an algebraic linear ordering 
that is isomorphic to the ordering $(L,\lex)$ where $L$ is a dcfl
over the ordered alphabet $A = \{a_0 < \ldots < a_{n-1}\}$. Then 
let $B = \{b_0,\ldots,b_{n-1}\}$ be ordered by $b_{n-1} < \ldots < b_0$,
and let $h: A^* \to B^*$ denote the homomorphism with $h(a_i) = b_i$,
for all $i \in [n]$. Then $h(L)$ is a dcfl and $P^*$ is isomorphic
to $(h(L),\lex)$. \eop

\section{Prefix and Scattered  Grammars}
\label{sec-grammars} 

In our proof of the fact that the Hausdorff rank of every scattered 
algebraic linear ordering is less than $\omega^\omega$ we will  make 
use of a corresponding result regarding the lexicographic 
ordering of languages generated by certain context-free grammars.
For all unexplained (but standard) notions regarding 
context-free grammars and languages we refer to \cite{HopcroftUllman}. 

Below we will consider context-free grammars  $G = (N,\{0, 1\},P,S)$
with set of nonterminals $N$, terminal alphabet $\{0, 1\}$, 
productions $P$ and start symbol $S$. We will assume that either 
$\L(G)$, the language generated by $G$, 
 is nonempty and $G$ 
contains no useless nonterminals, or $N = \{S\}$ and $P$ is empty.
For each $p \in (N \cup \{0,1\})^*$, we let $\L(p)$ denote the language
$$\L(p) = \{ w \in \{0,1\}^* : p \Rightarrow^* w\}$$
so that, in particular, $\L(G) = \L(S)$. 

The nonterminals of a context-free grammar $G = (N,\{0, 1\},P,S)$
may 
be classified into strong components in the usual way. 
We recall the necessary definitions. 

\begin{deff}
Suppose $X,Y$ are nonterminals.
Write
\begin{eqnarray*}
 Y &\preceq & X 
\end{eqnarray*}
if there is a derivation $X \Rightarrow^* p  Y  q $ 
for some $p $ and $ q $ in $(N \cup \{0,1\})^* $. Define
\begin{eqnarray*}
  X \approx Y &\iff & X \preceq Y \text{ and  }  Y \preceq X.
\end{eqnarray*}
\end{deff}
The relation $\preceq$ is a preorder on the nonterminals, and
induces a partial order on the equivalence classes 
\begin{eqnarray*}
  [X] &:=& \{Y: X \approx Y\} ,
\end{eqnarray*}
where $[Y]\leq [X]$ if $Y \preceq X$.
\begin{deff}
The \textbf{height} of a nonterminal $X$ is the number
of equivalence classes
 $[Y]$ strictly below $[X]$ in this ordering.    
\end{deff}

We note some elementary properties of height.

\begin{prop}
  \label{height prop}
Suppose $X,Y$ are nonterminals.
  \begin{itemize}
\item If $X\approx Y$, then $X$ and $Y$ have the same height.
  \item If the nonterminal $X$ has height $h$ and 
$Y \preceq X$, then $Y$ has height 
at most $h$.
\item If both $X,Y$ have height $h$ and if
$Y \preceq X$, then $X \approx Y$.
  \end{itemize}
\eop 
\end{prop}

In the proof of  Theorem \ref{main thm}, 
we make use of the following easy lemma.

\begin{lem}
\label{approx lem}
  Suppose $X \Rightarrow^* up$ is a (leftmost) derivation, where
$u \neq \epsilon \in \zos$ and $p \in (N \cup \{0,1\})^*$. If the nonterminal
$Y$ occurs in $p$ and $X \approx Y$, then there is a (leftmost) derivation
\begin{eqnarray*}
  X & \Rightarrow ^*& u v X q,
\end{eqnarray*}
for some $v \in \zos$ and $q \in (N \cup \{0,1\})^*$. \eop 
\end{lem}

Now we provide the definition of prefix grammars and scattered grammars that will
play a crucial role.

\begin{deff}
We call a context-free grammar $G$ a \textbf{prefix grammar}
if for each nonterminal $X$, $\L(X)$ is a prefix language. 
A \textbf{scattered grammar} is a prefix grammar $G$ such
that $(\L(G),\lex)$ is a scattered linear ordering. 
\end{deff}


In the definition of scattered grammars, we only required that the 
lexicographic ordering of the language generated from the start symbol
is scattered. As shown by the next result, it follows that the  
lexicographic ordering of the language generated from any nonterminal,
and in fact any word possibly containing both nonterminals and terminals, 
is scattered.

\begin{prop}
\label{prop-inherit}
If $G$ is a scattered grammar, then for each $p \in (N\cup \{0,1\})^*$, 
$(\L(p),\lex)$ is a scattered linear ordering. 
\end{prop}
{\sl Proof.} 
We may assume $\L(G) \neq \emptyset  $.  
For any $p = u_0X_1 u_1 \ldots u_{n-1}X_nu_n$, where each $X_i$ is 
a nonterminal, and each $u_i \in \zos$, the linear ordering $(\L(p),\lex)$ 
is isomorphic to the cartesian product
$$\L(X_n,\lex) \x \cdots \x  \L(X_1,\lex),$$
(note the reverse order), since each language $\L(X_i)$ is prefix.
Thus, since the cartesian product of a finite number of scattered linear orderings 
is scattered, by Corollary~\ref{cor-product0}, it suffices to prove that $(\L(X),\lex)$ is scattered 
for each nonterminal $X$. 

But for any nonterminal $X$ there exist words $u,v \in \{0,1\}^*$ with $S \Rightarrow^* uXv$, and 
if $u$ and $v$ are such words, then, as above, the linear ordering $(\L(X),\lex)$ is isomorphic to 
$(u\L(X)v,\lex)$, since $\L(X)$ is a prefix language. Moreover, there is an order embedding 
of $(u\L(X)v,\lex)$ into $(\L(S),\lex)$, which is a scattered linear ordering by assumption.
Thus, by Lemma~\ref{lem: ordrank2},  $(\L(X),\lex)$ is also scattered. \eop 

The next lemma gives a condition on a context-free grammar $G$ which
implies that $(\L(G),\lex)$ is not scattered.

\begin{lem}
Suppose that $G$ is a context-free grammar. Suppose that $X \in N$ and 
$u_i,v_i \in \{0,1\}^*$ with $X\Rightarrow^* u_iXv_i$, $i = 0,1,2$ and $u_0 <_s u_1 <_s u_2$.
Then $(\L(X),\lex)$ is not scattered. 
\end{lem}

{\sl Proof.}
Let $w$ denote a word in $\L(X)$. Define 
$$L_0 = \{ u_{i_1}\ldots u_{i_k}u_1 w v_1v_{i_k}\ldots v_{i_1} : i_1\ldots i_k \in \{0,2\}^*,\ k \geq 0\}.$$
Clearly, $L_0 \subseteq \L(X)$. Also, if $i_1\ldots i_k1 <_s j_1\ldots j_m1$, then 
$$ u_{i_1}\ldots u_{i_k}u_1 w v_1v_{i_k}\ldots v_{i_1} <_s u_{j_1}\ldots u_{j_m}u_1 w v_1v_{j_m}\ldots v_{j_1}. $$
This shows that $(\{0,2\}^*1, <_s)$ is isomorphic to $(L_0,<_s)$, which is in turn
isomorphic to $(L_0,\lex)$. But  it is easy to see that
$(\{0,2\}^*1, <_s)$ is isomorphic to the ordering of the rationals, 
so that $(L_0,\lex)$ is not scattered. \eop

We use the above lemma together with Proposition~\ref{prop-inherit} to prove:

\begin{prop}
\label{prop-prefix2}
Suppose that $G$ is a scattered grammar and $X$ is a nonterminal. If $X \Rightarrow^* uXp$ and 
$X \Rightarrow^* vXq$
where $u,v \in \{0,1\}^*$ and $p,q \in (N \cup \{0,1\})^*$, then either $u \leq_p v$ or $v \leq_p u$. 
\end{prop}
{\sl Proof.}
Let $u',v' \in \{0,1\}^*$ with $p \Rightarrow^* u'$ and $q \Rightarrow^* v'$. We have
the following derivations:
\begin{eqnarray*}
X &\Rightarrow^* uuXu'u'\\
X &\Rightarrow^* uvXv'u'\\
X &\Rightarrow^* vvXv'v'
\end{eqnarray*}
If neither $u$ is a prefix of $v$ nor $v$ is a prefix of $u$, then $u<_s v$ or $v<_s u$. 
In the first case, $uu <_s uv <_s vv$, while in the second $vv <_s vu <_s uu$, and thus by
the previous lemma, $(\L(X),\lex)$ is not scattered. 
This contradicts Proposition~\ref{prop-inherit}. \eop

\textit{In the rest of this section, we will assume that $G$ is a 
scattered grammar and 
for each nonterminal $X$, $\L(X)$ 
contains at least two words}. 

It follows that $G$ is $\epsilon$-free, 
i.e., there 
exists no production of the form $X \to \epsilon$. Since we may easily eliminate all chain productions 
$X \to Y$, where $X,Y$ are nonterminals, we will also assume that $G$ is free of chain productions. 

\begin{prop}
There exist no nonterminal $X$ and words $p,q,r \in (N \cup \{0,1\})^*$ 
with $X \Rightarrow^* pXqXr$.
\end{prop}

{\sl Proof.} Assume that $X \Rightarrow^* pXqXr$. Let $u,v \in \{0,1\}^*$ 
with $ p \Rightarrow^* u$ and $q \Rightarrow^* v$. Moreover,
let $x,y \in \L(X)$ be distinct words. We have that 
$X \Rightarrow^* uxvXr$ and $X \Rightarrow^* uyvXr$. Now  $x <_s y$ or $y<_s x$
and thus $uxv <_s uyv$ or $uyv <_s uxv$, contradicting Proposition~\ref{prop-prefix2}. \eop 

A nonterminal $X$ is \textit{recursive} if there is a derivation
$X \Rightarrow^+ pXq$, for some $p,q \in (V \cup \{0,1\})^*$, and \textit{left recursive} 
is there is a derivation $X \Rightarrow^+ Xq$ for some $q \in (V \cup \{0,1\})^*$.
 
\begin{prop}
$G$ is left-recursion free. 
\end{prop} 

{\sl Proof.} Assume that $X \Rightarrow^+ Xp$. Since $G$ is $\epsilon$-free 
and free of chain productions, we have that $p \neq \epsilon$
and there exists some nonempty word $u \in \L(p)$. Let $v \in \L(X)$. Since $vu$ is also in $\L(X)$, we conclude 
that $\L(X)$ is not a prefix language, contradiction. \eop 

Recall that a nonempty word $u \in\{0,1\}^*$ is \emph{primitive} if it cannot be written in the 
form $u=v^k$ for any $v \in \{0,1\}^*$ and integer $k \geq 2$. It is known, 
cf. \cite{Lothaire}, Proposition 1.3.1,  that each
nonempty word $u \in \{0,1\}^*$ can be written in a unique way as $u = v^k$,
where $v$ is primitive, called the \emph{primitive root} of $u$.

\begin{prop}
\label{prop-u0}
For every recursive nonterminal $X$ there is a primitive word $u_0 \in \{0,1\}^+$ such that 
whenever $X \Rightarrow^+ uXp$
for some $u \in \{0,1\}^*$ and $p \in (N \cup \{0,1\})^*$, then $u = u_0^n$ for some $n \geq 1$.
\end{prop} 

{\sl Proof.} Suppose that $X \Rightarrow^+ uXp$, where $u \in \{0,1\}^*$ and $p \in (N \cup \{0,1\})^*$.
Since $G$ is left recursion free, we have $u \neq \epsilon$.
Let $u_0$ be the primitive root  of $u$. We will show that whenever $X \Rightarrow^+ vXq$
with $v \in \{0,1\}^+$ and $q \in (N \cup \{0,1\})^*$ then there is some integer $n$ with 
$u_0^n = v$, i.e., $u_0$ is also the primitive root of $v$.

So assume that $X \Rightarrow^+ vXq$. There exist integers $k,m\geq 1$ with $|u^k| = |v^m|$. 
Since $X \Rightarrow^+ u^kXp^k$ and $X \Rightarrow^+ v^m X q^m$ and since $|u^k| = |v^m|$,
we have that $u^k = v^m$, by Proposition~\ref{prop-prefix2}. Thus $v^m$ is a power of $u_0$, and since 
$u_0$ is primitive, $v$ also must be a power of $u_0$, see \cite{Lothaire}. \eop

Sometimes we will  write $u_0^X$ for $u_0$. 

\begin{deff}
Let $X$ be a recursive nonterminal, $u_0 =u_0^X$.  Then for each $n \geq 0$ and prefix $u1$ of  $u_0$
(so that $u0$ is \emph{not} a prefix of $u_0$), let
\begin{eqnarray*}
L(X,n,u1) &=& \{u_0^nu0v \in \{0,1\}^* : X \Rightarrow^* u_0^n u0v\}.
\end{eqnarray*}
Similarly, for each $n \geq 0$ and prefix $u0$ of  $u_0$, let 
\begin{eqnarray*}
R(X,n,u0) &=& \{u_0^nu1v \in \{0,1\}^* : X \Rightarrow^* u_0^n u1v\}.
\end{eqnarray*}
Moreover, let 
\begin{eqnarray*}
L(X,n) &=& \bigcup_{u1 \leq _p u_0} L(X,n,u1)
\end{eqnarray*}
\begin{eqnarray*}
R(X,n) &=& \bigcup_{u0 \leq _p u_0} R(X,n,u0).
\end{eqnarray*}
\end{deff}

\begin{prop}
\label{prop-decomp}
\begin{itemize}
\item If $n < m$, and $x \in L(X,n),\ y \in L(X,m)$, then $x<_s y$. 
\item If $n < m$, and $x \in R(X,n),\ y \in R(X,m)$, then $y<_s x$. 
\item For any $n,m$ and words $x \in L(X,n)$ and $y \in R(X,m)$ it holds that $x <_s y$. 
\item  Finally, $\L(X) = L \cup R$, where $L = \bigcup_n L(X,n)$ and 
 $R = \bigcup_n R(X,n)$. 
\end{itemize}
\end{prop}
{\sl Proof.} It is
easy to check the first three claims.

Consider now a left derivation $X \Rightarrow^* w$ of some word
$w \in \{0,1\}^*$. Let $k$ be large enough so that $u_0^k$ is longer than $w$. Since 
$\L(X)$ is a prefix language and contains a word of the form $u_0^kx$, it follows that 
$w<_s u_0^k$ or $u_0^k <_s w$. In the first case, we can write $w$ in the form 
$w = u_0^nu0v$ for some words $u,v$ such that $u1$ is a prefix of $u_0$,
so that $w \in L(X,n,u1)$.
In the second case, $w = u_0^nu1v$ for some words $u,v$ such that $u0$ is a prefix 
of $u_0$ and thus  $w \in R(X,n,u0)$. \eop 

\begin{cor}
\label{omega+omega* sum}
$ (\L(X),\lex) $ is isomorphic to the sum $L + R$, where $L,R$ are the generalized sums
  \begin{eqnarray}
\label{eq-L,R,1}
     L &=& \sum_{n \geq 0} (L(X,n), \lex) \\
\label{eq-L,R,2}
     R &=& \sum_{n \leq 0} (R(X,-n), \lex).
  \end{eqnarray}
\end{cor}
{\sl Proof.\/ } 
Immediate, from the previous proposition. \eop

The following theorem is one of our main results.

\begin{thm}
\label{main thm}
Suppose that $X$ is a nonterminal in $G$ of height $h$. Then 
$$r((\L(X),\lex)) \leq \omega^h + 1.$$
\end{thm}

{\sl Proof.} 
Suppose the height of $X$ is 0.  
Note that if $Y\preceq X$, then $X \approx Y$ by Proposition
\ref{height prop}, so that if $Y \neq X$ then $X$ is
recursive. Thus, if $X$ is not recursive, since $X$ has height 0, 
$\L(X)$ is finite, and
\begin{eqnarray*}
  r((\L(X),\lex)) & \leq & 1.
\end{eqnarray*}
If $X$ is recursive, $\L(X)$ is isomorphic to $L + R$ where 
$L$ and $R$ are defined in (\ref{eq-L,R,1}) and  (\ref{eq-L,R,2}).
We claim that for any $n \geq 0$, each language $L(X,n)$ and $R(X,n)$ is finite.  Indeed,
to prove $L(X,n)$ is finite,
it is enough to prove that if $u1 \leq_p u_0$, then $L(X,n, u1)$ is finite, since
$ L(X,n)=\bigcup_{u1 \leq_p u_0} L(X,n,u1)$, where $u$ ranges over all words in $\{0,1\}^*$ 
such that $u1$ is a prefix of $u_0 = u_0^X$.
Thus, suppose
\begin{eqnarray*}
  X & \Rightarrow^* u_0^n u0 p
\end{eqnarray*}
is a leftmost derivation, where $p \in (N \cup \{0,1\} )^*$.  We claim that
$p$ cannot contain a nonterminal.  Otherwise, if the nonterminal $Y$ occurs
in $p$, $X \approx Y$, as noted above, so by Lemma
\ref{approx lem},
there is a derivation $X \Rightarrow^* u_0^n u0 w X q$,
contradicting Proposition \ref{prop-u0}.  
Thus, $p$ is a word in $\zos$.  There is a finite number of leftmost such derivations
with $p \in \zos$, showing $L(X,n,u1)$ is finite.  

The argument showing $R(X,n)$ is finite is the same. 
Thus,  $(\L(X),\lex)$ is isomorphic to $\omega + \omega^*$, so that
$r((\L(X),\lex)) \leq 2 = \omega^0 + 1.$

Assume now that $h > 0$.

{\em Case 1.} $X$ is not recursive. Then whenever $X \to p$ is a production, 
any nonterminal occurring in $p$ is of height less than $h$,
by Proposition \ref{height prop}.
By the induction hypothesis and Corollary~\ref{cor-product}, 
it follows that the Hausdorff rank of each $(\L(p),\lex)$ such that $X \to p$
is a production is less than $\omega^h$. Since $\L(X)$ is a finite union of 
such languages $\L(p)$, by Corollary~\ref{cor-union}, the Hausdorff rank of $(\L(X),\lex)$ 
is less than $\omega^h$.

{\em Case 2.} $X$ is recursive. Then, by
Corollary \ref{omega+omega* sum},
 $(\L(X),\lex)$ is isomorphic to the
sum $L+R$, where $L$ and $R$ are defined in (\ref{eq-L,R,1}) and (\ref{eq-L,R,2}).
Below we will show that the rank of each $(L(X,n),\lex)$ is 
less than $\omega^h$. The rank of each $(R(X,-n),\lex)$ 
is also less than $\omega^h$, as may be shown in the same way,
so that, by Corollary~\ref{cor-product}, the rank of $(\L(X),\lex)$ is at most $\omega^h + 1$.

Now for each fixed $n$, $L(X,n)$ is itself a finite union
$ \bigcup_{u1 \leq_p u_0} L(X,n,u1)$, where $u$ ranges over all words in $\{0,1\}^*$ 
such that $u1$ is a prefix of $u_0 = u_0^X$. Since $G$ is $\epsilon$-free and free of 
left recursion, for each fixed prefix $u1$, $L(X,n,u1)$ is a \emph{finite} union 
of languages of the form $\L(u_0^nu0p)$, where there is a left derivation $X \Rightarrow^+ u_0^nu0p$.
 It follows that the nonterminals occurring in any such $p$ 
have height less than $h$. 
Indeed, if a nonterminal $Y$ with height $h$
occurs in $p$, then $X\approx Y$ and by Lemma
\ref{approx lem},
there is a derivation $X \Rightarrow^+ u_0^nu0vXr$,
where $v$ is a terminal word. But $ u_0^nu0v$ is not a power of $u_0$,
contracting Proposition
\ref{prop-u0}.
Thus, by Corollary~\ref{cor-product},
we obtain  that  the rank of any $(\L(u_0^nu0p),\lex)$ is less than $\omega^h$. Thus, by 
Corollary~\ref{cor-union}, it follows that 
 $r((L(X,n),\lex)) < \omega^h$, for all $n$. \eop 

\begin{cor}
\label{bound-cor}
  If $G$ is a scattered grammar, $r((\L(G),\lex)) < \omega^\omega $. 
\end{cor}
{\sl Proof.\/ } If the number of nonterminals in $G$ is $n$, then
the height of any nonterminal is less than $n$. \eop

\section{From Algebraic Trees to Prefix Grammars}
\label{sec-translation}

Consider a system of equations
\begin{eqnarray}
\label{a system}
F_i(x_0,\ldots,x_{n_i-1}) &=& t_i(x_0,\ldots,x_{n_i-1}), \quad i=1,\ldots,m. 
\end{eqnarray}
where each $t_i$ is a term over the ranked alphabet $\Sigma \cup \F$
in the variables $x_0,\ldots,x_{n_i-1}$. 
We assume that $F_1$ is the principal function variable and that $n_1 = 0$. 
Each component of the least solution is an algebraic tree. 

 Let
$T$ be a finite or infinite tree
in $T_{ \Sigma \cup \F}^\omega(V)$. 
The labeled frontier languages of trees in 
$T_{ \Sigma \cup \F}^\omega(V)$ are defined as above.


(As mentioned above, it is known that a locally finite tree $T$ is 
algebraic iff  $\Lfr(T)$ is a deterministic context-free language.)
We recall from \cite{BEord} the following result: 

\begin{thm}
\label{alg to prefix}
When $T \in T_\Sigma^\omega$ is an algebraic tree, $\Lfr(T)$ can be generated by a 
prefix grammar. 
\end{thm}

Let $(T=T_1, T_2,\ldots, T_m)$ denote the least solution of the 
system (\ref{a system}). For the reader's convenience we recall  from \cite{BEord}
the construction of a prefix grammar generating $\Lfr(T)$. 

We will define a grammar whose nonterminals $N$ consist of the 
letters $F_i$, together with all ordered pairs $(F_i,j)$ where 
$i = 1,\ldots,m$, $j \in [n_i]$.
The alphabet   $\Gamma$ of terminal letters 
is $\Gamma = \Sigma_0 \cup \{(\gamma,j): \gamma  \in (\Sigma \cup \F)_k,\ j \in [k]\}$.
The grammar is designed to have the following properties.

\textbf{Claim:}
\textit{For any word $u$, $T_i(u) = x_j$ iff $(F_i,j) \Rightarrow^* \widehat{u}$.
And for any word $u$, $T_i(u) \in \Sigma_0$ iff $F_i \Rightarrow^* \widehat{u}T_i(u)$.
Moreover, any terminal word derivable from $(F_i,j)$ is of the form $\widehat{u}$, 
and any terminal word derivable from $F_i$ is of the form $\widehat{u}T_i(u)$ 
for some $u \in \dom(T_i)$.}

The grammar generating $\Lfr(T)$ is: $G_L= (N,\Gamma,P,F_1)$, 
where the set $P$ of productions is defined below. 
If $t_1,\ldots, t_m$ are the terms
on the right side of (\ref{a system}) above, 
then the productions are: 
\begin{itemize} 
\item 
$$(F_i,j) \to \widehat{u}$$
where $u \in \dom(t_i)$ and $t_i(u) = x_j$,
\item 
$$F_i \to \widehat{u}t_i(u)$$
where  $u \in \dom(t_i)$ and  $t_i(u) \in \Sigma_0 \cup \F$.
\end{itemize}
The proof of the fact that this grammar is a prefix grammar 
generating the language $\Lfr(T)$ can be found in \cite{BEord}. 
\eop 

\begin{cor}
\label{alg to prefix grammar}
If $(Q,<) $ is an algebraic linear ordering, there is a prefix grammar
$G'$ with $(Q,<)$ isomorphic to $ (\L(G'),\lex)$. 
\end{cor}

\begin{remark}
\label{remark-poly}
The above constructions can be carried out in polynomial time. 
Thus, each recursion scheme (over $\Delta$) defining an 
algebraic linear ordering can be transformed in polynomial 
time into a prefix grammar defining the same linear ordering. 
\end{remark}

\section{Completing the Proof}
\label{sec-completing}

To finish the argument,
we apply Corollary \ref{alg to prefix grammar}.

\begin{thm}
\label{alg to grammar}
\label{main result1}
The Hausdorff rank of any scattered algebraic linear order
is less than $\omega ^\omega $.
\end{thm}
{\sl Proof.\/ } 
Indeed, any scattered algebraic linear ordering is isomorphic
to the leaf ordering of an algebraic tree, by
Proposition \ref{prop-frontier} above. 
By Corollary 
\ref{alg to prefix grammar}, there is a prefix grammar $G$ such that this leaf
ordering is isomorphic to $(\L(G),\lex)$. It follows that
$G$ is a scattered grammar.
By Corollary
\ref{bound-cor}, $r((\L(G),\lex))<\omega^\omega $. 
\eop 

\begin{cor} [\cite{BEord}]
  \label{main result2} 
The algebraic ordinals are precisely those less than $\ooo$.
\end{cor}
{\sl Proof.\/ } 
Every ordinal less than $\ooo$ is algebraic, by 
Corollary \ref{cor: ub on ords}.
We need to show there
are no more algebraic ordinals. But,
if $(Q,<)$ is an algebraic well-ordering of order type $\alpha $, then
$r(\alpha )<\omega^\omega $, by Corollary \ref{main result1}.  But then $\alpha < \ooo$,
by Corollary~\ref{cor: ordrank2}. 
\eop 

\section{Conclusion and Open Problems}
\label{sec-conclusion}

A hierarchy  of recursion schemes was introduced in \cite{Indermark}, see also 
\cite{Damm,Damm82,Gallier,Ong,HagueOng}, and many others. 
Here, we dealt with level $0$ (regular schemes) and level $1$ 
(algebraic or first-order schemes) of the hierarchy. 
In Theorem~\ref{main result1}, we have shown that every scattered linear ordering 
definable by a level $1$ scheme has Hausdorff rank less than $\omega^\omega$,
whereas it has been known that the Hausdorff rank of any scattered linear 
ordering definable by a recursion scheme of level $0$ is less than $\omega$. 
We conjecture that for each $n$, the Hausdorff rank of 
any scattered linear ordering definable by 
a level $n$ scheme is less than $\Uparrow(\omega,n+1)$, a tower of $n+1$ 
$\omega$'s. If that conjecture is true, then it follows 
that an ordinal is definable by a level $n$ scheme if and only if it is less than 
$\Uparrow(\omega,n+2)$, and thus an ordinal is definable in the hierarchy if and only if 
it is less than $\epsilon_0$. (See also \cite{Braud}, where it is shown that 
any ordinal less than $\epsilon_0$ is definable in the hierarchy.) 

In ordinal analysis of logical theories, the strength of a theory is measured 
by ordinals. For example, the proof theoretic ordinal of Peano arithmetic is 
$\epsilon_0$. Here we have a similar phenomenon: we measure the strength 
of recursive definitions by ordinals, and we conjecture that the ordinals definable 
are exactly those less than $\epsilon_0$. We also conjecture that the Hausdorff rank 
of any scattered linear ordering definable in the hierarchy of recursion schemes 
is less than $\epsilon_0$. The same may hold for the Caucal's pushdown hierarchy,
\cite{Caucal}.

Finally, we mention two more open problems.

\textbf{Problem} 
Characterize the context-free well orderings and scattered linear 
orderings. 

\textbf{Problem}
Is it decidable for two algebraic linear orderings (each specified
by a recursion scheme) whether they are isomorphic? 

Put in other way, the question is whether it is decidable 
for two dcfl's equipped with the lexicographic ordering whether 
they are isomorphic.


\end{document}